\documentclass[showpacs,preprint,aps]{revtex4} 

\usepackage{graphicx} 

\begin{document} 


\title{Non-stochastic behavior of atomic surface diffusion on Cu(111) at all 
	temperatures} 
\author{J. Ferr\'{o}n} 
\affiliation{Grupo de F\'{\i}sica INTEC-FIQ, CONICET, Universidad Nacional 
	del Litoral, 3000-Santa Fe, Argentina.} 
\author{L. G\'omez} 
\affiliation{Facultad de Ciencias Exactas, Ingenier\'{\i}a y Agrimensura, 
	Instituto de F\'{\i}sica Rosario, 2000-Rosario, Argentina.} 
\author{J.J. de Miguel and R. Miranda} 
\affiliation{Departamento de F\'{\i}sica de la Materia Condensada and 
	Instituto de Ciencia de Materiales "Nicol\'{a}s Cabrera"; \\ 
	Universidad Aut\'{o}noma de Madrid, Cantoblanco, 28049-Madrid, 
	Spain.} 

\date{\today} 

\begin{abstract} 
Atomic diffusion is usually understood as a succession of random, independent 
displacements of an adatom over the surface's potential energy landscape. 
Nevertheless, an analysis of Molecular Dynamics simulations of self-diffusion 
on Cu(111) demonstrates the existence of different types of correlations in the 
atomic jumps at all temperatures. Thus, the atomic displacements cannot be 
correctly described in terms of a random walk model. This fact has a profound 
impact on the determination and interpretation of diffusion coefficients. 
\end{abstract} 


\maketitle 

Surface diffusion is a most influential process at the atomic scale, lying 
at the core of many relevant fields. In catalysis and surface chemistry the 
ability of the reactants to come together and/or reach the active surface 
sites controls the reaction rates \cite{catalysis}. Likewise, in epitaxial 
growth the adatom mobility decisively influences the nucleation probability 
and the average distance and size of the objects -islands- formed \cite{tringides}; 
kinetic limitations result in the accumulation of roughness as growth proceeds 
\cite{zhang}. 

Despite the many efforts devoted to studying this phenomenon, our current 
understanding is far from being complete \cite{barth}. The continuous 
advancements in both theory and experiments are unveiling a rich phenomenology 
unsuspected till now; even for the simplest self-diffusion case, new basic 
mechanisms are still being discovered. First it was site exchange, detected 
by means of FIM experiments \cite{kellogg} and theoretically demonstrated 
shortly afterwards \cite{feibelman}. Later, evidence was found on the existence 
of "long jumps", i.e., atomic displacements spanning several lattice constants 
\cite{ehrlich,linderoth}, lifting the restriction to nearest-neighbor hops. The 
last finding so far is subsurface diffusion, in which adatoms move below a loosely 
bound overlayer deposited on a given surface \cite{ferron,neugebauer}. 

Traditionally, diffusion was described as a random walk in which the adatoms 
occasionally acquire sufficient energy to jump from one lattice site to a 
nearby one where they attain again thermal equilibrium with the substrate; 
they would thus spend most of the time residing at the local minima of the 
potential energy surface and only a short fraction of it crossing the barriers. 
With these assumptions and applying Transition State Theory (TST) one gets for 
the diffusion coefficient \cite{glasstone}: 
\begin{equation}
D = D_0(T) \exp (-E_m/k_B T), 
\label{D} 
\end{equation} 
where $E_m$ is the difference between the saddle point and the minimum 
energy configurations along the most probable path. $D_0$ is slightly 
temperature-dependent, but it can be safely taken as a constant 
\cite{kurpick1,kurpick2}. 

Using a different approach, the diffusion process can be described in terms 
of the so-called hopping frequency: 
\begin{equation} 
\gamma = \gamma_0 \exp (-E_m/k_B T). 
\label{gamma} 
\end{equation} 
$D$ and $\gamma$ are geometrically related \cite{kurpick2} only if the 
diffusion events are stochastic. When the thermal energy is higher than the 
diffusion barrier ($k_B T \ge E_m$) the adatoms move rather freely over the 
surface and the definition of a hopping frequency loses sense. In this regime, 
the diffusion coefficient is expected to follow a linear behavior, $D = k_B 
T/m G$, where $G$ is an appropriate friction coefficient for the unconfined 
Brownian motion of the adatom \cite{einstein}. It is generally accepted that 
at temperatures of the order of $E_m / 2 k_B$ most of the assumptions of TST 
fail, diffusion cannot be represented by a random walk anymore and long, 
correlated jumps become more and more important \cite{sanders}. Hence at room 
temperature (RT, 300 K), an $E_m$ not smaller than 50 meV is required to 
satisfy the above condition. For compact metallic faces such as Cu(111) the 
activation energy for monomer diffusion is clearly smaller \cite{stoltze,karimi}, 
and therefore correlated displacements are to be expected well below RT. 

In this work, we demonstrate the existence of other correlated atomic movements 
even at low temperature, when single hops predominate, thus questioning many 
assumptions accepted so far. We draw our conclusions from analyzing the outcome 
of Molecular Dynamics (MD) simulations in which we follow the displacement of 
an adatom over a fully relaxed surface at different temperatures. Our MD code 
\cite{dynamo} uses interatomic potentials based on the Embedded Atom Model 
(EAM) \cite{foiles}. The sample was a slab of 14 layers with 270 atoms in each, 
and vacuum on both sides; periodic boundary conditions were used in all 
directions. The three bottom layers were frozen to simulate the bulk. The 
evolution of a single adatom at the upper surface was followed for different 
temperatures. In order to obtain reliable statistics, the simulation was 
extended for up to 10 ns for the lower temperatures. 

Surface diffusion is quantitatively described by means of the coefficient $D$, 
which can be calculated from the adatom displacements through the Einstein 
relation in 2 dimensions: 
\begin{equation} 
\sigma^2(t) = \frac{1}{N} \left [\sum_{i=1}^N (x_i(t) - x_0)^2 + (y_i(t)-y_0)^2 
	\right ] = 4\ D\ t 
\end{equation} 
where $x_i(t)$, $y_i(t)$ are the surface coordinates at time $t$ for $N$ 
different initial conditions. As stated above, the diffusion coefficient may 
also be determined from the hopping frequency, $\gamma$, as $D = \gamma l^2/4$ 
for (111) faces \cite{kurpick2}, $l$ being the hop length (1.47 \AA\ for 
Cu(111)). 

Within the framework of TST both descriptions are equivalent. However, in the MD 
simulation the mean square displacement $\sigma^2(t)$ of the diffusing particle 
can be obtained independently of $\gamma$. The latter is calculated as 
the number of successful jumps divided by the total elapsed time. In Fig. 
\ref{Dfig} we depict in the usual form of an Arrhenius plot the values of the 
diffusion coefficient $D_\sigma$ obtained from the mean square displacement, 
and that found from the analysis of the hopping frequencies ($D_\gamma$). It 
is noteworthy that $D_\sigma$ clearly deviates from the Arrhenius law at high 
temperatures. A similar phenomenon was already observed by Kallinteris {\it et 
al.} \cite{kallinteris}, who ascribed this behavior to the onset of a new 
activated mechanism, namely the diffusion along the [110] direction by means 
of double jumps \cite{linderoth}. Besides, a fit to $D_\gamma$, which displays 
a more Arrhenius-like behavior, yields an activation energy of $30 \pm 1$ meV 
(solid line in Fig. \ref{Dfig}), much larger than the theoretical predictions 
\cite{stoltze,karimi}. 

Fig. \ref{trajectories} shows some representative atomic trajectories at 
different temperatures. In general they consist of displacements, which 
frequently extend up to several lattice sites, separated by periods of 
vibration within a single surface cell; the relative abundance of each type of 
event depends on the temperature. Similar trajectories have been reported 
previously, resulting from simulations with other sets of interatomic potentials 
\cite{montalenti} and also in experiments \cite{senft}. Evidently, adatom 
diffusion at high temperature can hardly be considered a random phenomenon; any 
thermally activated process, characterized by an attempt frequency (pre-exponential 
factor) and an energy barrier, should have the same occurrence probability for 
all equivalent paths. Quite on the contrary, clearly deterministic trajectories 
can be observed in Fig. \ref{trajectories}(c); the path followed by the Cu adatom 
is reminiscent of surface channeling, where an energetic diffusing particle is 
steered along its trajectory by the interaction potential and its interactions 
with the surrounding atoms. The enhanced diffusivity at high temperature, above 
the expected Arrhenius behavior, is thus caused by these correlated movements. On 
the other hand, at low temperatures random processes based on single jumps 
separated by long stays at a given adsorption well seem to be dominant and no 
preferred trajectories are evident. 

Further insight into the applicability of TST can be gained from a detailed 
study of the atomic trajectories. We split the surface into two sets of cells, 
corresponding to the fcc and hcp sites, and measure the time spent by the atom 
at every adsorption well by detecting the transitions from one cell to another 
one nearby. Fig. \ref{histograms} shows histograms depicting the statistical 
distribution of such residence times for different temperatures. The appearance 
of large peaks at very short times is an evidence for the lack of randomness in 
the diffusion process. The average residence time for each substrate temperature 
is marked by the vertical line in the graphs of Fig. \ref{histograms}. The shift 
to lower times (i.e., higher jump frequencies) is clearly due to the disappearance 
of several types of processes, most significantly the recrossings, with their 
longer associated times, and not to a larger jump probability, as would be 
for a thermally activated process. 

The stochastic nature of the diffusive process can be analyzed in full detail 
by sorting out the different kinds of jumps: all those that can be shown to be 
statistically non-independent will therefore be considered correlated. 
In Fig. \ref{histograms}, the solid circles take into account all 
the jumps detected in our simulations. A striking feature of this figure is the 
appearance of a double peak at short residence times for the lower temperatures. 
The first one corresponds to what we call {\it ballistic jumps}: rapid crossings 
making up the long displacements in which the adatom traverses several surface 
cells in a single impulse. This is obviously the shortest residence time observed 
and is related to the average velocity of the diffusing adatoms. We shall call 
this elementary time interval $\tau_0$; its magnitude decreases with increasing 
substrate temperature, from 0.6 ps at 100 K down to 0.38 ps at 650 K, reflecting 
the higher kinetic energy of the adatoms. 

The curves marked with open squares in Fig.~\ref{histograms} depict the times 
associated with recrossing events \cite{sanders}, that is, two consecutive jumps 
that bring the adatom back to its former position. At 100 K this kind of 
processes completely accounts for the second peak in the general distribution. 
Significantly, this peak is centered at a time close to $2 \tau_0$. Below 185 
K yet another clear peak can be seen at about $3\tau_0$. As expected, this 
triple transition time appears after two frustrated jump attempts. Quite 
surprisingly, almost none of these hops is a recrossing, as demonstrated by 
the statistics in Figs. \ref{histograms}a and \ref{histograms}b; 1/3 of them 
would be expected if the process were truly random. We shall thus call this 
kind of jumps ``double-bounces'' for brevity. 

Our statistical analysis of diffusion jumps is presented in Fig. \ref{correl}. 
First, it is remarkable the large number of ballistic jumps observed even at 
low temperature; needless to say, they absolutely dominate the scene above 300 
K. At low temperature the recrossings are more frequent than expected (1/3 of 
the total) for a random process. This is probably due to the geometric 
arrangement of atoms in the (111) face: The adatom jumping through the saddle 
point is directed in a collision trajectory toward the third atom in the 
threefold cell. If the energy is not enough to set the channeling effect in, 
the atom is backscattered  preferently in a recrossing trajectory. Our results 
show that these processes occur before the adatom becomes thermalized again. 
Neither of them can be considered stochastic; rather, this behavior hints toward 
some participation of the substrate atoms. As for the double-bounces, their 
topological analysis is also shown in Fig. \ref{correl}b: despite the rise at 
about 200 K, the percentage of them that are recrossings never reaches 1/3, 
implying that the choice of directions is also non-random. We thus conclude that 
at least three oscillations within an adsorption well, or a residence time $\ge 
4\tau_0$ is required to ensure the statistical independence of diffusion 
hops. 

To summarize our findings, we find evidence in our simulations showing that 
most of the atomic jumps in surface diffusion are biased, i.e., the process is 
not stochastic: the correlated jumps exceed 95\% of the total above 500 K and 
even at 100 K amount to more than 50\%. The kind of correlations, however, is 
not the same, and they have different effects on the determination of diffusion 
coefficients. Below $\sim$200 K recrossing events dominate, they compute as 
jumps but do not produce an effective displacement and therefore cause an 
overestimation of $D_\gamma$ in this temperature range. On the contrary, at high 
temperatures the predominance of long jumps in the forward direction results in 
the longer net displacements responsible for the deviation in the values of 
$D_\sigma$. Clearly then, the simple, intuitive connection between the diffusion 
coefficients and the activation energy for nearest-neighbor atomic hops can only 
be established if all correlated jumps are excluded. This can easily be done 
with our simulation results: recalculating $D_\gamma$ in this way yields the 
results depicted in Fig. \ref{Dprime}, with a preexponential $D'_0 = 0.44$ 
\AA$^2$/ps and an activation energy $E_m = 22 \pm 1$ meV which, although still 
higher than the calculated static barriers, approaches much more closely their 
values. In any case, these results must be taken with care since at high 
temperatures the statistics is poor due to the small number of uncorrelated 
jumps. 

The physical picture that emerges from this work is the following: due to 
thermal fluctuations the adatom gains energy and momentum. The former allows it 
to overcome the diffusion barriers, while the latter establishes a privileged 
movement direction \cite{jacobsen}. Due to the symmetry of the (111) face, the 
diffusing adatom is propelled in a collision trajectory toward the surface atom 
situated directly across the nearest-neighbor three-fold site. If the adatom 
has enough energy, the surface potential steers its trajectory along the [110] 
direction, initiating the ballistic movement. On the other hand, if its kinetic 
energy is not high enough the adatom will be backscattered by the surface atom, 
and may perform a recrossing. In any case, the truly statistically independent 
jumps seem to be a minority at all temperatures of interest. Obviously, the key 
point here is the rate at which the adatom transfers its energy to the substrate 
to become thermalized again. Further work is in progress on this subject. All 
these effects influence the relationship between the atoms displacements and 
the hopping frequencies, and therefore also the experimental determination of 
diffusion barriers \cite{zhang2}. We expect that our findings may shed some light 
onto these controversies. 

\begin{acknowledgments} 
This work has been financed by the CICyT through project BFM2001-0174 and by 
CONICET through PIP 2553/99. J.F. thanks Fundaci\'{o}n Antorchas for financial 
support.  
\end{acknowledgments}

\newpage 

\begin{figure}[htb]
\caption{Diffusion coefficient for Cu atoms on Cu(111), calculated either by 
	measuring the atomic displacement as a function of time ($D_\sigma$, solid 
	squares) or from the hopping frequency ($D_\gamma$, open circles). The 
	solid line is an Arrhenius fit to $D_\gamma$, with $D_0 = 1.56 \times 10^{-4} 
	{\rm cm^2~s^{-1}}$ and $E_m = 30~\pm~1$ meV.} 
\label{Dfig} 
\end{figure}

\begin{figure}[!htb]
\caption{Some typical examples of atomic trajectories of Cu adatoms 
	self-diffusing on Cu(111) at diverse temperatures. Notice the different 
	sizes of the regions exposed.}
\label{trajectories} 
\end{figure}

\begin{figure}[!htb] 
\caption{Statistical distribution of residence times of the diffusing Cu monomers 
	as a function of substrate temperature. Solid circles: all kinds of jumps; 
	open squares: recrossings. The vertical lines mark the average residence 
	time at each particular temperature; at 100 K the corresponding value of 
	10.6 ps falls outside the graph range. The inset within the last graph 
	illustrates the definition of the surface cells with a typical ballistic 
	trajectory and the determination of residence times: the solid circles are 
	surface atoms, while the smaller grey and open dots mark the hcp and fcc 
	sites, respectively.}
\label{histograms} 
\end{figure}

\begin{figure}[htb]
\caption{Statistical analysis of correlations in atomic jumps. 
	(a) Percentage with respect to the total of: ballistic or long jumps (open 
	circles), recrossings (squares) and ``double-bounces'' (diamonds). The dotted 
	line is the 1/3 mark, the expected probability of recrossings for purely 
	random hops. 
	(b) Percentage of ``double-bounces'' (jumps after a residence time $3\tau_0$) 
	that return to the previous adsorption site. The total fraction of correlated 
	jumps, after our analysis, is given by the solid triangles in (a); even at 
	100 K this figure exceeds 50\%, reaching nearly 100\% at the higher 
	temperatures.} 
\label{correl} 
\end{figure}

\begin{figure}[htb]
\caption{Diffusion coefficient $D'_\gamma$ (filled diamonds) obtained from the 
	hopping frequency $\gamma$ after excluding all correlated jumps. The solid 
	line is an Arrhenius fit to these data, yielding $D_0 = 4.44 \times 10^{-3} 
	{\rm cm^2~s^{-1}}$ and $E_m = 22 \pm 1$ meV. The open circles are the 
	uncorrected $D\gamma$ values from Fig. \ref{Dfig}, shown for comparison.} 
\label{Dprime} 
\end{figure}

\end{document}